\documentclass[12pt,preprint]{aastex}

\shorttitle{Rotational Modulation of M/L Dwarfs}
\shortauthors{Lane et al.}

\begin{document}


\title{Rotational Modulation of M/L Dwarfs due to Magnetic Spots}


\author{C. Lane\altaffilmark{1}, G. Hallinan\altaffilmark{1}, R. T. Zavala\altaffilmark{2},  R. F. Butler\altaffilmark{1}, R. P. Boyle\altaffilmark{3}, S. Bourke\altaffilmark{1},\\  A. Antonova\altaffilmark{4}, J. G. Doyle\altaffilmark{4}, F. J. Vrba\altaffilmark{2}}


\and

\author{A. Golden\altaffilmark{1}}


\altaffiltext{1}{Centre for Astronomy, National University of Ireland, Galway, Ireland; c.lane2@nuigalway.ie, gregg@it.nuigalway.ie, ray.butler@nuigalway.ie, stephen@it.nuigalway.ie,  agolden@it.nuigalway.ie}
\altaffiltext{2}{United States Naval Observatory, Flagstaff Station, 10391 West Naval Observatory Road, Flagstaff, AZ 86001; bzavala@nofs.navy.mil, fjv@nofs.navy.mil}
\altaffiltext{3}{Vatican Observatory Research Group, Steward Observatory, University of Arizona, Tucson, AZ 85721; boyle@ricci.as.arizona.edu}
\altaffiltext{4}{Armagh Observatory, College Hill, Armagh BT619DG, N. Ireland; tan@arm.ac.uk, jgd@arm.ac.uk}


\begin{abstract}
We find periodic I-band variability in two ultracool dwarfs, \object{TVLM 513-46546} and \object{2MASS J00361617+1821104}, on either side of the M/L dwarf boundary. Both of these targets are short-period radio transients, with the detected I-band periods matching those found at radio wavelengths (P=1.96 hr for TVLM 513-46546, and P=3 hr for 2MASS J00361617+1821104). We attribute the detected I-band periodicities to the periods of rotation of the dwarfs, supported by radius estimates and measured $v$ sin $i$ values for the objects. Based on the detected period of rotation of TVLM 513-46546 (M9) in the I-band, along with confirmation of strong magnetic fields from recent radio observations, we argue for magnetically induced spots as the cause of this periodic variability. The I-band rotational modulation of L3.5 dwarf 2MASS J00361617+1821104 appeared to vary in amplitude with time. We conclude that the most likely cause of the I-band variability for this object is magnetic spots, possibly coupled with time-evolving features such as dust clouds.
\end{abstract}





\section{Introduction}
The term ultracool dwarf encompasses very low mass stars and brown dwarfs, covering spectral types late-M, L and T. As the temperature decreases, spectral type moves from M to L, with the characteristic TiO and VO bands of M dwarf spectra becoming replaced by broad neutral alkali and iron hydride lines in L dwarfs at roughly 2000 K (Kirkpatrick et al., 2000). Although the change in spectral features in progressing between spectral types M and L is well characterised, the processes occurring in their cool, neutral atmospheres are not fully understood. Also uncharacterised are any differences in the physical nature or extent of surface features.

One of the main reasons for photometric variability studies of late-M and L dwarfs is to probe the physical nature of their atmospheres. Ultracool dwarf variability is generally attributed to two sources: the presence of magnetic spots, or dust clouds (Mart\'{i}n et al., 2001; Gelino, 2002 (G02); Rockenfeller et al., 2006 (R06)). Surface features such as magnetic spots or dust clouds may cause optical modulation as the object rotates, and in certain cases allow a measurement of the period of rotation of a dwarf. For either of these scenarios to be accepted as the dominant source of photometric variability for a particular ultracool dwarf, they must explain the type of variability observed, whether it is aperiodic or periodic. It is sometimes possible to constrain the probable cause of variability for an object, by comparing multiwavelength photometry to photometric signatures generated using atmospheric models such as those of Allard et al. (2001). This method has recently been used to indicate magnetically induced cool spots as the most likely cause of periodic variability (P=3.65 $^+_-$ 0.1 hr) in the M9V dwarf 2MASS 1707+64 (R06). 

There are indications of changes in the physical nature or the extent of surface features when moving from spectral type M to L. Time-resolved photometry of a statistical sample of M and L dwarfs in directly comparable datasets suggests that variability is more common in L dwarfs than late-M dwarfs (Bailer-Jones and Mundt 2001 (BJM01); R06). Other photometric studies have demonstrated a high level of L dwarf variability, with a large-scale L dwarf survey reporting variability in 7 out of 18 L dwarfs at the 95.4$\%$ confidence level, with a further five targets showing variability at confidence levels of $\sim$ 80$\%$ (G02).

Another characteristic of L dwarf photometric variability is a lack of stability in both periodicity and amplitude. Aperiodic variability is the dominant type of L dwarf variability found in the surveys by BJM01 and G02. Two of the seven variables in the latter survey, 2MASS 0746+20AB and 2MASS 1300+19, displayed significant peaks in a CLEAN periodogram (Roberts et al., 1987), but did not display persistent periodicity throughout the dataset. Koen (2006) reported a 2.4 hr periodicity for 2MASS 0605-22342, which persisted over 3 days, but decreased significantly in amplitude throughout the observations. This apparent paucity of stable periodic variability may be linked to spectral type, although there has been no long-timescale monitoring of ultracool dwarf variability to establish this statistically.

It has been suggested that rapid evolution of atmospheric features such as dust clouds is the most likely cause of observed aperiodic variability in L dwarfs (G02). However, it may be possible that magnetic spots on L dwarfs also produce significant periodic modulation, but that evolving features such as dust obscure the underlying period. One theory which may explain the characteristics of L dwarf variability is the masking hypothesis, proposed by BJM01, which suggests that if the time scale of evolving surface features is shorter than the period of rotation of an object, the periodic modulation of its lightcurve will be inhibited. In order to assess the validity of such theories of variability for a particular object, constraints on the value of its period of rotation from $v$ sin $i$ measurements and radius estimates are extremely useful. Limits on the period of rotation determine whether a period detected in photometric timeseries is rotational, and allow an evaluation of the stability of the photometric periodicity in relation to it.

Recent Very Large Array (VLA) observations of the M9 dwarf TVLM 513-46546 (hereafter TVLM 513) provided evidence of a 1.96 hr stable periodicity (Hallinan et al., 2006, 2007 (H06 \& H07)). This periodicity was found to be consistent with a coherent radiation mechanism generated via the electron cyclotron maser instability operating in the low plasma density regions above the magnetic poles of the dwarf, requiring TVLM 513 to possess extremely strong ($\sim$ kG) magnetic fields. Such strong fields were later confirmed for other ultracool dwarfs by Reiners and Basri (2007). H06 also argues that the L3.5 dwarf  2MASS J00361617+1821104 (hereafter 2MASS J0036+18) requires the same mechanism to explain the properties of its radio emission (Berger, 2005 (B05)), which are almost identical to those of TVLM 513. 

The extremely strong magnetic fields of TVLM 513, and most likely for 2MASS J0036+18, might be expected to produce the necessary temperature gradient to create magnetic spots, thus causing photometric modulation. Consequently, we decided to photometrically monitor these dwarfs on either side of the M/L boundary for I-band variability. 

\section{Data Acquisition \& Observations}
\subsection{Supporting Radio Data}
The supporting radio observations discussed in this paper involve $\sim$ 10 hours of VLA observations of TVLM 513 at a frequency of 8.4 GHz on 2006 May 20th and $\sim$ 10 hours at a frequency of 4.9 GHz on 2006 May 21st. These observations and the accompanying results are detailed in H07. The 4.9 GHz radio data of 2MASS J0036+18 is that of B05, and was obtained from the VLA archives (project AB1052).\\

\subsection{I-band Observing Strategy}
Choices made with regard to the optical observations are as follows: (1) the TVLM 513 observations were chosen to coincide with the nights of additional VLA observations (H07), (2) due to the very red optical colours of these dwarfs, the I-band was chosen to monitor these objects for variability, (3) high time-sampling was performed for sensitivity to the rotational modulation of these rapid rotators.

\subsection{TVLM 513 Observations}
The four-night photometric campaign of TVLM 513 ($m_I \sim 15.09$) was carried out between 2006 May 18th and 21st, UT. Observations were performed on the United States Naval Observatory (USNO) 1m reflector, Flagstaff, Arizona from May 18th - 20th. A Tektronix 2048 $\times$ 2048 CCD was used, with pixel scale of 0.68\arcsec/pixel giving a field of view of 23\arcmin $\times$ 23\arcmin. An imaging cadence of $\sim$ 3 mins was achieved over the $\sim$ 6 hours of observations on each night. Additional observations were carried out on May 20th and 21st at the 1.8m Vatican Advanced Technology Telescope (VATT) at Mount Graham, Arizona. A back-illuminated Loral CCD was windowed to a 512 $\times$ 512 pixel frame (field of view=1.7\arcmin $\times$ 1.7\arcmin) to reduce readout time. This CCD showed evidence of fringing above the sky noise level in the I-band, so frames were dithered to allow removal of fringing during the data reduction.  An imaging cadence of $\sim$ 1 min was obtained for the May 21st dataset. Unfortunately, due to telescope tracking problems, the data obtained on May 20th were of poor quality and were not used in the analysis.

\subsection{2MASS J0036+18 Observations}
Photometric observations of 2MASS J0036+18 ($m_I \sim 16.05$) were carried out on 2006 September 19th UT, using the USNO 1.55m reflector at Flagstaff, Arizona. The detector used was the Tek2K CCD, providing an image scale of 0.33\arcsec/pixel and an 11$\arcmin$  $\times$ 11$\arcmin$ field of view. Over the $\sim$ 7.5-hour observation of 2MASS J0036+18, well-exposed images were obtained every $\sim$ 8 mins.

\section{Data Reduction}
\subsection{Basic Reduction}
Science frames of TVLM 513 and 2MASS J0036+18 were processed using standard IRAF\footnote[1]{Image Reduction and Analysis Facility.}/PyRAF data reduction techniques and the USNO on-site reduction pipeline. Bias subtraction and flat-fielding were carried out on all frames. The 2MASS J0036+18 data were also linearized using the Tek2K linearity curve. The fringing pattern of the VATT CCD was found to be stable, and its removal was carried out by producing a fringe-correction frame from the median of a large number of dithered, reduced science frames. 

\subsection{$I$-band Differential Photometry}
Differential photometry was carried out on all datasets in order to achieve photometric precision of the order of milli-magnitudes through the reduction of atmospheric effects. Aperture photometry was carried out on the targets and a selection of reference stars, chosen on the basis of their stability, linearity, isolation on the frame, and having a similar magnitude to the target. Apertures were chosen to provide the highest signal to noise for the target. Standard differential photometry techniques were used, similar to those outlined in BJM01, and summarized below.  Time series of relative magnitudes ($m_{rel}$) were calculated as follows: If F$_i$ is the instrumental flux for reference star i, with range of reference stars i=1...M, then the relative flux of the r$th$ frame (F$_r$) is:
\begin{equation}
F_r = \frac{1}{M}\sum_{i}^{M} F_i
\end{equation}

The $m_{rel}$ values were calculated as follows, from corresponding target and relative magnitudes, m$_t$ and m$_r$, where F$_t$ is the measured flux of the target: \begin{math}m_{rel} = m_t - m_r = 2.5 log_{10} (F_r/F_t)\end{math}.

\subsection{Error Estimation}
Assuming the reference stars were well selected, the standard deviations of their non-variable differential lightcurves may be used to make an empirical estimate of the total photometric errors for a target of a certain magnitude (BJM01). This approach may be used to estimate both formal and ``informal" errors, such as those from flat-fielding and fringing, which may be difficult to evaluate accurately. The advantage of this strategy is that it models the expected error in a target due to noise alone, which may be difficult to otherwise isolate in a target with intrinsic variability. The photometric errors in the individual reference star lightcurves were calculated by plotting their formal photometric errors from IRAF ($\sigma_{IRAF}$) against the standard deviations of their lightcurves ($\sigma_{rms}$). The free parameters of the first order polynomial fit to this plot ($a$ and $b$) were then used to give the errors in the i$th$ reference star lightcurve ($\delta m_{i}$): \begin{math}\delta m_{i} = a + b \cdot \sigma_{IRAF}\end{math}. The error in the relative magnitude ($\delta m_{rel}$) was then calculated using the magnitude errors in the target $\delta m_{t}$, and the i$th$ reference star, $\delta m_{i}$ (i=1...M), as demonstrated by BJM01: 
 
 \begin{equation}
( \delta m_{rel})^2 = (\delta m_{t})^2 +( \frac{1}{MF_r})^2 \sum_{i}^{M} F_i^2 (\delta m_{i})^2
 \end{equation}

\subsection{Period Determination and Significance of Variability}
Periodograms or power spectra were used to search for periodic variability. The Lomb-Scargle Periodogram (Lomb, 1976; Scargle, 1982) was calculated for the I-band data of TVLM 513 and 2MASS J0036+18. The power spectra or periodograms were searched for significant peaks, which signify periodic variability. When a significant peak was found, the data were phase folded to this period. These phase folded lightcurves were then visually inspected as a final test - the scatter of the lightcurves were found to be lowest when folded to the periods detected in the periodograms, and began to show increased scatter when folded at other periods. The measure of variability amplitude used is $\sigma_{rel}$, the standard deviation of the target lightcurve. However, the significance of variability in a lightcurve is not simply the ratio $\sigma_{rel}$/$\overline{\delta m_{rel}}$, as for a large number of points, statistically significant variability may be detected even if  $\sigma_{rel}$ is only slightly larger than $\overline{\delta m_{rel}}$ (see BJM01). 





\section{Results}
Low amplitude, quasi-sinusoidal variability was detected in the I-band photometry for both objects, with periods identical to those found at radio wavelengths (P=1.96 hr for TVLM 513, and P=3 hr for 2MASS J0036+18); see Fig 1 of this paper and Fig 2 of H07. These periods were used with radius estimates (Dahn et al., 2002) to calculate rotational velocities of $\sim$ 60 km s$^{-1}$ for TVLM 513 and $\sim$ 37 km s$^{-1}$ for 2MASS J0036+18. The calculated velocities are consistent with  $v$ sin $i$ measurements of $\sim$ 60 km s$^{-1}$ for TVLM 513 (Basri, 2001) and 36 $^+_-$ 2.7 km s$^{-1}$ for 2MASS J0036+18 (Zapaterio Osorio, 2006), indicating a high inclination angle, $i$. Based on these results, we attribute the I-band periods to the periods of rotation, with rotation axes perpendicular to the line of sight. 

The detected rotational modulation in the TVLM 513 I-band data ($\sigma_{rel}$=0.0076, $\overline{\delta m_{rel}}$=0.0054), along with confirmation of strong magnetic fields in the radio data suggests that magnetic spots are the most probable cause of variability. Additionally, this periodic modulation appeared to be sustained over the four nights of the observing campaign, being detected in periodograms of the optical data on all nights between May 18th and 20th (USNO) and in the May 21st data (VATT). There is a high degree of correlation in the lightcurves from both telescopes, despite instrumental differences (see Fig 2). 

The rotational modulation found for 2MASS J0036+18 is interesting ($\sigma_{rel}$=0.015, $\overline{\delta m_{rel}}$=0.0089),  as the amplitude of the modulation apparently changes within the timescale of the observation, a phenomenon which is not reflected in the lightcurves of reference stars or of a field star of similar magnitude (see Fig 2). Just less than three cycles are visible in the 2MASS J0036+18 lightcurve, with the amplitude of the first `peak' most closely matching the third (Fig 2). Due to the short observation, the Lomb Scargle periodogram assumes that this behaviour will repeat ad infinitum, and thus a second peak is produced in the periodogram at a frequency of half the fundamental (Fig 1). As the 3 hr periodicity provides a rotational velocity consistent with $v$ sin $i$ measurements, and there is no physical reason for the 6 hr periodicity, the variation in the peak to peak amplitude is most likely due to some underlying aperiodic variability. If this object is indeed an analogue of TVLM 513 as argued by H06, it must have strong magnetic fields. Therefore, we suggest that magnetic spots are the most likely source of the 3 hr rotational modulation, while rapidly evolving features may cause the apparent change in amplitude of the variability. Such features might include dust clouds or magnetic spots which are changing in size or temperature.






\section{Discussion \& Conclusions}
We have detected periodic modulation in the I-band photometry of TVLM 513 and 2MASS J0036+18, which we have determined to be rotational using radius estimates and $v$ sin $i$ measurements. Radio observations have provided confirmation of strong magnetic fields for TVLM 513 and there is evidence for their existence on 2MASS J0036+18. Based on the optical and radio results, we argue that magnetic spots are the cause of the variability of both these objects.

In the case of 2MASS J0036+18 (L3.5), this is contrary to expectations that dust is the cause of all L dwarf variability. It has been suggested that the fractional ionization of the photospheric and atmospheric plasma of ultracool dwarfs is too low to sustain magnetic spot structures (G02, Mohanty et al., 2002). However, if the magnetic field is large-scale, as has been confirmed to be the case for TVLM 513 (H07), it may produce magnetic spots. A plausible explanation for the 2MASS J0036+18 result is that the amplitude of the underlying rotational signal due to magnetic spots is itself modulated by evolving features such as dust clouds. However, long-timescale observations are necessary to rule out random noise effects and to investigate the possible influence of evolving features.

It should be noted that our observations were designed to be sensitive to the rotational modulation of rapid rotators, with high temporal resolution photometry obtained for both objects. This strategy was found to be successful, and our next challenge is to monitor 2MASS J0036+18 to investigate timescales for possible evolving features by isolating their effects from the underlying rotational modulation. 




\acknowledgments
This work is supported by Science Foundation Ireland under its Research Frontiers Programme, the Higher Education Authority's Programme for Research in Third Level Institutions, and the Irish Research Council for Science, Engineering and Technology. Armagh Observatory is grant-aided by the N. Ireland Dept. of Culture, Arts \& Leisure. We also thank Hugh Harris and our referee for their suggestions, which have greatly improved the manuscript.

\clearpage

\begin{figure}
\plotone{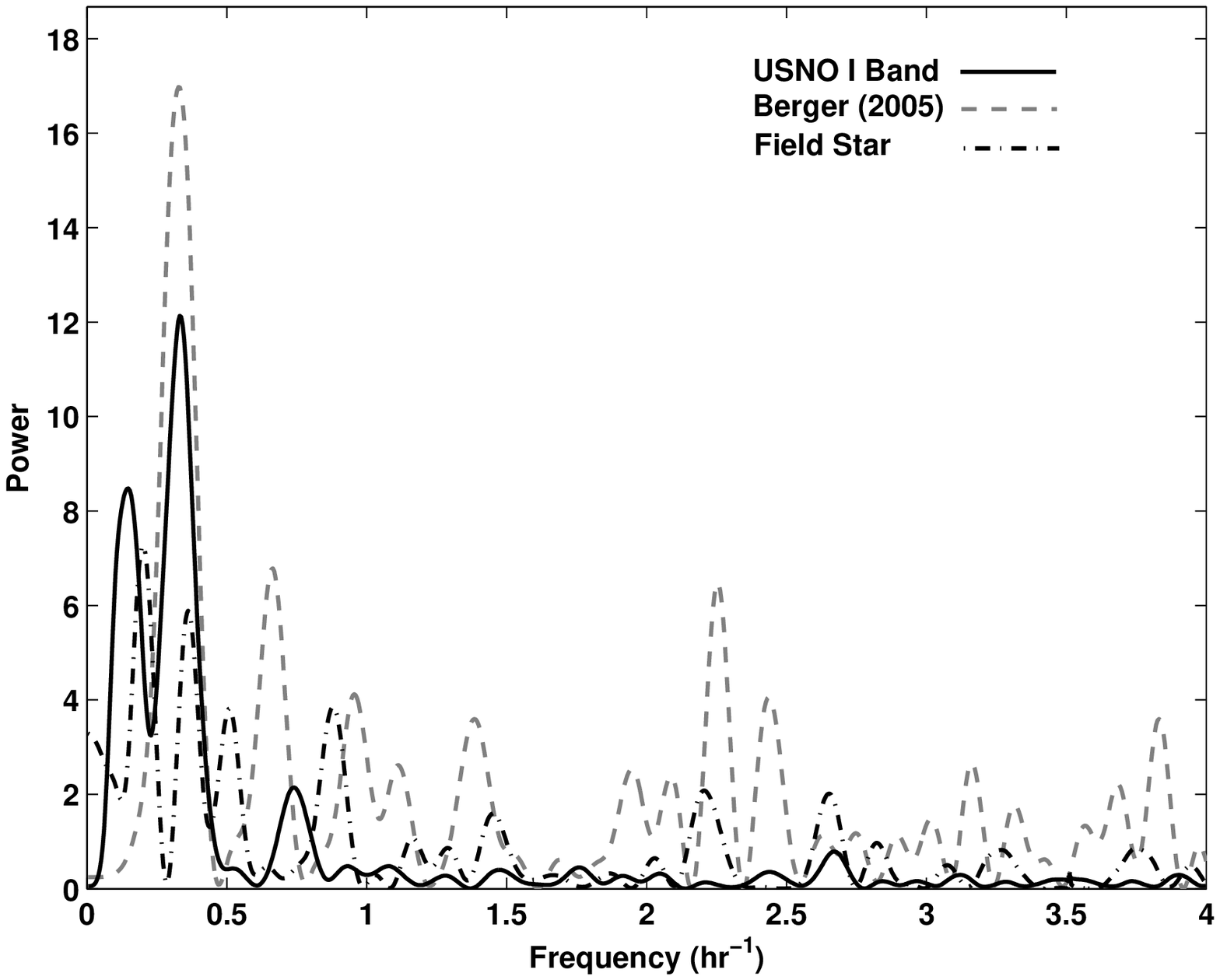}
\caption{Lomb-Scargle Periodograms of raw radio and optical data. I-band photometry of 2MASS J0036+18  and a field star of similar magnitude, obtained on USNO 1.55m, Sept 19th, 2006 and 4.9 GHz total intensity radio emission, B05.}
\end{figure}

\begin{figure}
\plotone{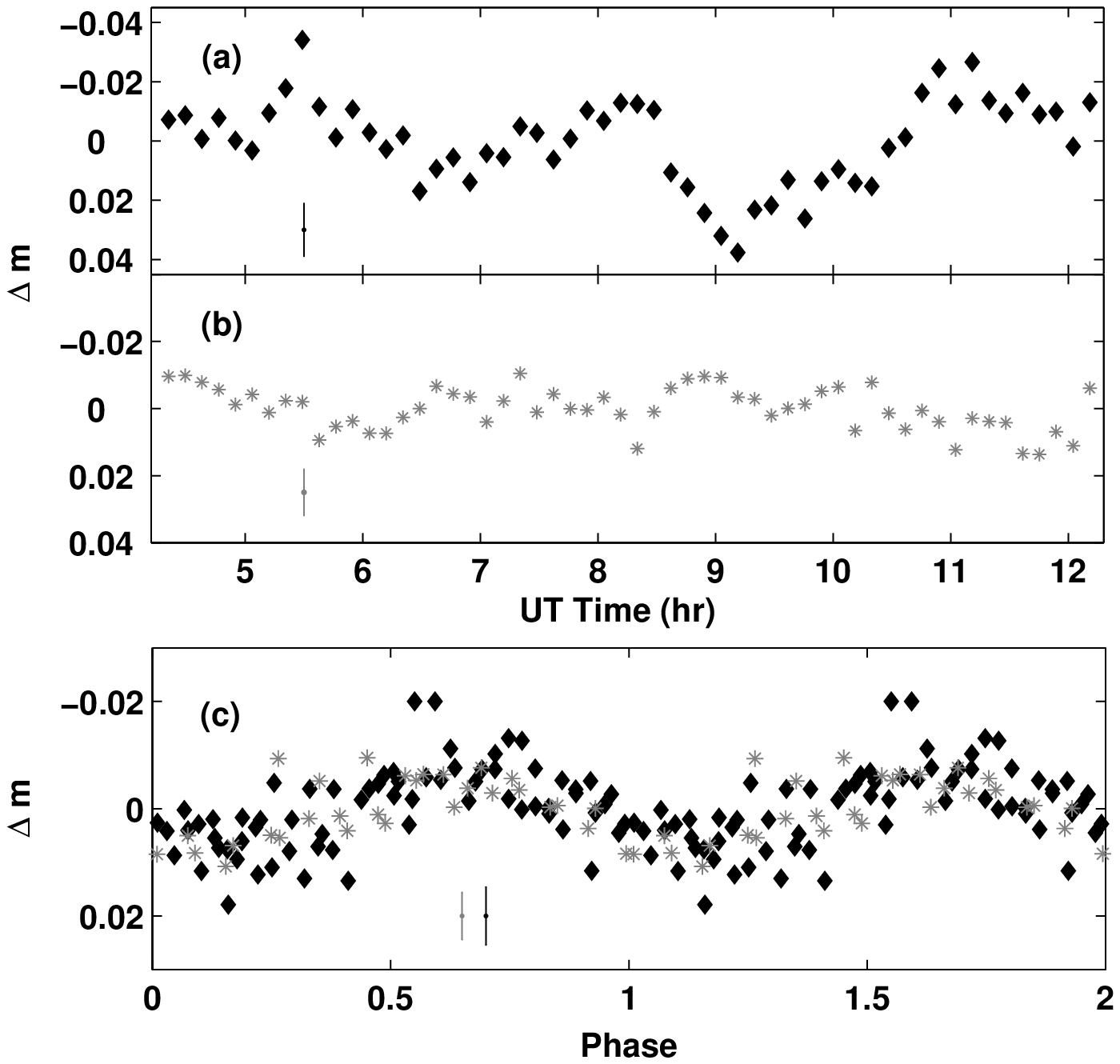}
\caption{I-band lightcurves of 2MASS J0036+18 and TVLM 513 (a) USNO 1.55m raw 2MASS J0036+18 lightcurve, September 19th UT, 2006, measured using 9 reference stars. (b) USNO 1.55m raw lightcurve of a field star of similar magnitude to 2MASS J0036+18, using same reference stars. (c) Binned, phase-folded lightcurves of TVLM 513 data, obtained at VATT 1.8m, May 21 2006 (black) and USNO 1m, May 18 2006 (grey), both measured using two reference stars. The error bars are taken as $\overline{\delta m_{rel}}$, the average expected error in the lightcurves.}
\end{figure}


\begin{thebibliography}{}
\bibitem[Allard(2001)]{al01} Allard, F., Hauschildt, P. H., Alexander, D. R., Tamanai, A., \& Schweitzer, A. 2001, \apj, 556, 357
\bibitem[Basri(2001)]{bas01} Basri, G. 2001, ASP Conf. Ser. 223, Cool Stars, Stellar Systems and the Sun: 11th 
Cambridge Workshop, ed. R. J. Garcia Lopez, R. Rebolo, \& M. R. Zapaterio Osorio (San Francisco: ASP), 261 
\bibitem[Bailer-Jones and Mundt]{bjm01} Bailer-Jones, C. A. L., \& Mundt, R. 2001, \aap, 367, 218 (BJM01)
\bibitem[Berger(2005)]{ber05} Berger, E., et al. 2005, \apj, 627, 960 (B05)
\bibitem[Dahn(2002)]{dahn02} Dahn, C. C, Harris, H. C., Vrba, F. J., Guetter, H. H., Canzian, B., Henden, A. A., Levine, S. E., Luginbuhl, C. B., Monet, A. K. B., Monet, D. G., Pier, J. R., Stone, R. C., Walker, R. L., Burgasser, A. J., Gizis, J. E., Kirkpatrick, J. D., Liebert, J., Reid, I. N., 2002, \aj, 124, 1170
\bibitem[Gelino(2002)]{gel02} Gelino, C. R., Marley, M. S., Holtzman, J. A., Ackerman, A. S., \& Lodders, K. 2002, \apj, 577, 433 (G02)
\bibitem[Hallinan(2006)]{hal06}Hallinan, G., Antonova, A, Doyle, J. G., Bourke, S., Brisken, W. F. \& Golden, A. 2006, \apj, 653, 690 (H06)
\bibitem[Hallinan(2007)]{hal07}Hallinan, G., Bourke, S., Lane, C., Antonova, A., Zavala, R.T.,  Brisken, W.F., Boyle, R.P., Vrba, F.J., Doyle, J.G.  \& A. Golden, 2007, ApJ, L663, 25 (H07)
\bibitem[Kirkpatrick(2000)]{kirk00}Kirkpatrick, J. D., Reid, I. N., Liebert, J., Gizis, J. E., Burgasser, A. J., Monet, D. G., Dahn, C. C.; Nelson, B., Williams, R. J., 2000, \aj,120, 447
\bibitem[Koen(2006)]{koen06} Koen, C., 2006, \mnras, 367, 1735
\bibitem[Lomb(1976)]{lomb76} Lomb, N. R. 1976, \apss, 39, 447
\bibitem[Mart\'{i}n(2001)]{martin01}Mart\'{i}n, E. L., Zapatero Osorio, M. R., \& Lehto, H. J., 2001, \apj, 557, 822
\bibitem[Mohanty(2002)]{moh02} Mohanty, S., Basri, G., Shu, F., Allard, F., \& Chabrier, G. 2002, \apj, 571, 469
\bibitem[Reiners(2007)]{reiners07} Reiners, A., \& Basri, G. 2007, \apj, 656, 1121
\bibitem[Roberts(1987)]{roberts87}Roberts, D. H., Lehar, J., \& Dreher, J. W. 1987, \aj, 93, 968
\bibitem[Rockenfeller(2002)]{rock06} Rockenfeller, B, Bailer-Jones, C.A.L, and Mundt, R. 2006, \aap, 448, 1111
\bibitem[Scargle(1982)]{scar82} Scargle, J. D., 1982, \apj, 263, 835
\bibitem[Zapatero Osorio(2006)]{zap06} Zapatero Osorio, M. R., Mart\'{i}n, E. L., Bouy, H., Tata, R., Deshpande, R. \& Wainscoat, R. 2006, \apj, 647, 1405

\end{thebibliography}
\end{document}